\documentclass{article}
\usepackage[a4paper, total={16cm, 24.7cm}]{geometry}

\usepackage{graphicx} 
\usepackage{float} 
\usepackage[utf8]{inputenc} 
\usepackage[T1]{fontenc} 
\usepackage{url} 
\usepackage[square,numbers]{natbib} 
\usepackage{authblk} 
\usepackage{array} 
\usepackage{multicol} 
\usepackage{multirow} 

\newcommand{\tsup}[1]{\textsuperscript{#1}}
\newcommand{\tsub}[1]{\textsubscript{#1}}

\title{Influence of Self-Absorption on Pulse Shape Discrimination\\ in Organic Glass Scintillators}
\author[1,2]{Lukasz Adamowski}
\author[1]{Martyna Grodzicka-Kobylka}
\author[1]{Tomasz Szczesniak}
\author[1]{Agnieszka Syntfeld-Ka\.zuch}
\author[1]{Lukasz Swiderski}
\author[2]{Adam Kisiel}
\affil[1]{National Centre for Nuclear Research (NCBJ), Otwock-Swierk, Poland}
\affil[2]{Faculty of Physics, Warsaw University of Technology, Warsaw, Poland}

\date{\today}

\begin{document}

\maketitle

\begin{abstract}
Organic glass scintillators are an interesting alternative to liquid scintillators, offering many advantageous characteristics with few drawbacks.
In this paper we investigate the influence of light self-absorption in the organic glass scintillator on its pulse shape discrimination capability.
With five scintillators of different heights but same diameter, we measure photoelectron yield and Figure of Merit in neutron-gamma discrimination.
The decrease of both values with increasing size is attributed to light self-absorption, while normalized Figure of Merit remains constant.
The choice of gates for charge comparison method is discussed.
We also use genetic algorithm to estimate decay times and intensities of fast, medium, and slow components of light pulse shapes measured with Bollinger-Thomas setup. We compare the results to \mbox{trans-stilbene} reference sample.
\end{abstract}

\section{Introduction} \label{sec:intro}
Organic glass scintillators (OGS) were developed by Patrick L. Feng and Joseph S. Carlson from Sandia National Laboratories \cite{Sandiawebsite_OGSdiscovery}. They have been granted patents for their work, including U.S. Patent 9,845,334 B1 for ``High-efficiency organic glass scintillators'' \cite{patent_US_9845334_B1} and U.S. Patent 10,508,233 B1 for ``Mixed compound organic glass scintillators'' \cite{patent_US_10508233_B1}.
Currently, OGS can be purchased from Sandia National Laboratories and Blueshift Optics, a company that has been granted a nonexclusive commercial license to manufacture and sell organic glass scintillators as well as perform research and develop them \cite{Sandiawebsite_BlueshiftOptics}. 
\par
OGS scintillators are considered among the best materials for discriminating between neutron and gamma radiation, with performance comparable to liquid scintillator EJ-309. 
Their advantages are that they are safer to use than liquid scintillators (which are extremely flammable and usually toxic), easier to produce and hence cheaper than \mbox{trans-stilbene} crystals, and brighter than plastic scintillators, while being more resistant to weather conditions and aging.
However, a notable drawback is their self-absorption of light, a property we began investigating in our previous research \cite{GrodzickaKobylka2023NIMA}.
\par
All scintillation materials capable of discriminating between neutron and gamma radiation exhibit a significant dependence of the Figure of Merit (FOM) and light output on the size of the scintillator itself. This is likely due to self-absorption of light within the scintillator material, a problem we investigated in our previous work \cite{GrodzickaKobylka2023NIMA} using 2\texttimes2 inch scintillator samples stacked on top of each other. In this article, we expand this research by utilizing new samples provided by Blueshift Optics and employing new methods to estimate optimal PSD conditions and decay times of pulse components, using the algorithm described in \cite{Adamowski2025pulses}.
All new samples have a diameter of 25.4 mm (1 inch) but varying lengths from 25~mm to 125~mm, which eliminated the need for stacking them (as was done in \cite{GrodzickaKobylka2023NIMA}) and the risk of light loss at interfaces. The proposed new offline analysis allows for much faster comparison of a large number of results, as well as rapid investigation of changes in neutron-gamma discrimination with various PSD parameters.

\section{Experimental details} \label{sec:experimental}

\begin{figure}[!htb]
    \centering
    \includegraphics[width=0.59\linewidth]{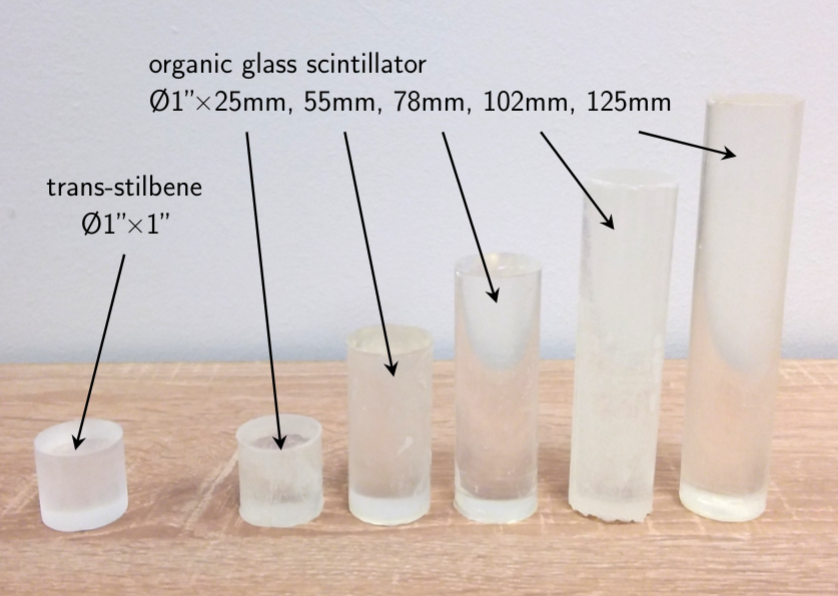}
    \caption{Photo of \mbox{trans-stilbene} (left) and five organic glass scintillators (right) used in this research.}
    \label{fig:photo}
\end{figure}

We characterized five organic glass scintillator (OGS) samples of different heights (ranging from 25 mm to 125 mm) provided by Blueshift Optics as well as a single 25.4 mm (1 inch) \mbox{trans-stilbene} crystal from Inrad Optics, which was used as a reference (Fig.~\ref{fig:photo}). All samples were cylindrical, with a diameter of 25.4 mm (1 inch).
\par
\begin{figure}[!htb]
    \centering
    \includegraphics[width=0.9\linewidth]{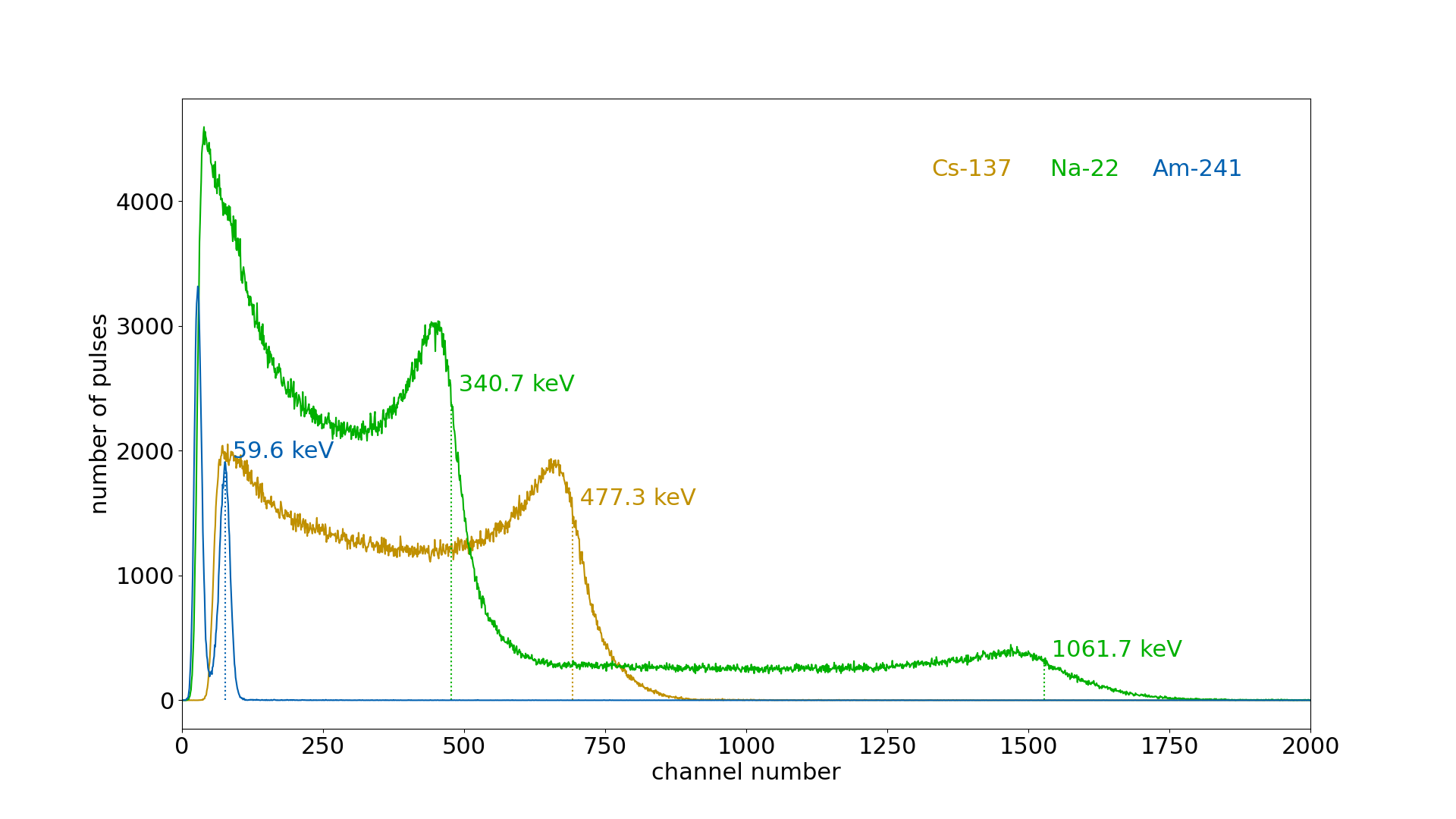}
    \caption{Energy spectra measured for 25-mm OGS with three calibration sources: \tsup{137}Cs, \tsup{22}Na and \tsup{241}Am. Three Compton edges of 1061.7, 477.3 and 340.7~keV, and the 59.6~keV full-energy peak are depicted.}
    \label{fig:calibration_spectra}
\end{figure}

We used a PuBe neutron source to test all samples' response to both neutron and gamma radiation. Fast neutrons interact with \mbox{hydrogen-rich} organic scintillators via recoil protons, while \mbox{gamma-rays} interact via electrons.
Only a small fraction of the particle's energy is deposited in the scintillator, and this fraction depends on the type of particle. Neutrons usually result in a much lower amount of energy being converted into light compared to \mbox{gamma-rays} of the same energy; therefore direct comparison between them is impossible.
In such cases, the keVee (keV electron equivalent) unit is used to describe deposited energy, where 1~keVee is defined as the energy of light deposited by a 1~keV fast electron \cite{Knoll2010}.
Calibration of energy deposited by gamma radiation was based on three gamma sources: \tsup{241}Am, \tsup{137}Cs, and \tsup{22}Na.
It was done independently in each of experimental setups, using sources appropriate for the setup and the energy range.
In small-scale low-density scintillators like the characterized samples no full-energy peaks were obtained for \tsup{137}Cs and \tsup{22}Na. Therefore, Compton edges were used for energy calibration: 477.3~keV in the case of \tsup{137}Cs, and 340.7~keV and 1061.7~keV in the case of \tsup{22}Na (Fig. \ref{fig:calibration_spectra}). Compton edges were estimated at 80\% height of the Compton maximum height \cite{Swiderskietal2010}. In the case of \tsup{241}Am, a full-energy peak of 59.6~keV was visible in the energy spectrum and was used for calibration (see Fig. \ref{fig:calibration_spectra}).
The calibration performed using gamma sources was also used as a reference for the energy deposited by fast neutrons.
\par

\subsection{Photoelectron yield measurement setup} \label{sec:experimental_yield}

We estimated photoelectron yield of all OGS samples using the Bertolaccini, et~al. method \cite{Bertolaccini1968} with \tsup{137}Cs gamma source. 
To maximize the light output samples were wrapped with white Teflon tape on all sides, except for those that were used to transfer light to a 2 inch Hamamatsu photomultiplier tube (PMT) model R6231-100. For the same reason we used silicone grease (Baysilone Ol M of various viscosity, from 100,000 mm\tsup2/s to 1,000,000 mm\tsup2/s) to couple the bare optical surfaces of the scintillators with the PMT's photocathode.
The scintillator and photomultiplier were covered with black adhesive tape and placed in a lightproof box.
For single photon spectrum measurements, the PMT without a sample was similarly protected from light and left in the dark for at least an hour to de-excite.
We used a precisely regulated high-voltage power supply to power the PMT via a voltage divider.
An ORTEC 672 Spectroscopy Amplifier with Canberra 2005E preamplifier and Tukan-8k-USB multichannel analyzer (MCA) were used to obtain both the \tsup{137}Cs gamma Compton continuum and single photon spectra.
The linearity of the amplifier was also measured with an external signal generator and taken into account.
\par

\begin{figure}[!htb]
    \centering
    \includegraphics[width=0.9\linewidth]{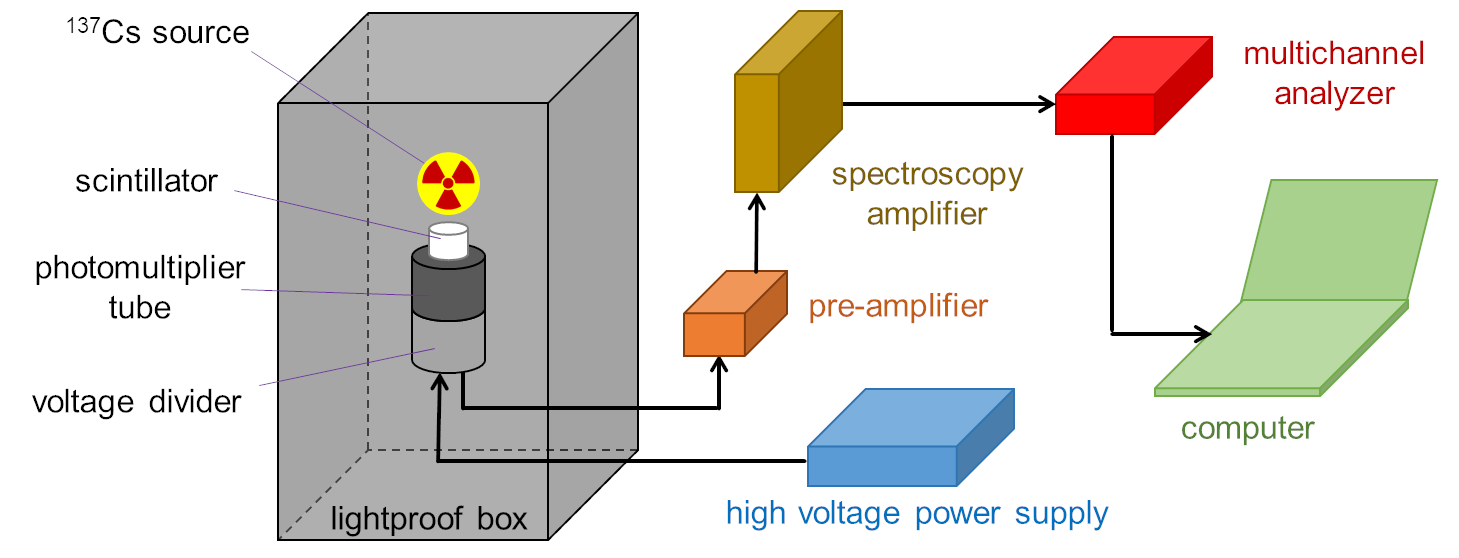}
    \caption{Experimental setup for photoelectron yield measurement.}
    \label{fig:experimental_setup_yield}
\end{figure}

\subsection{Neutron-gamma discrimination measurement setup} \label{sec:experimental_FOM}

\begin{figure}[!htb]
    \centering
    \includegraphics[width=0.9\linewidth]{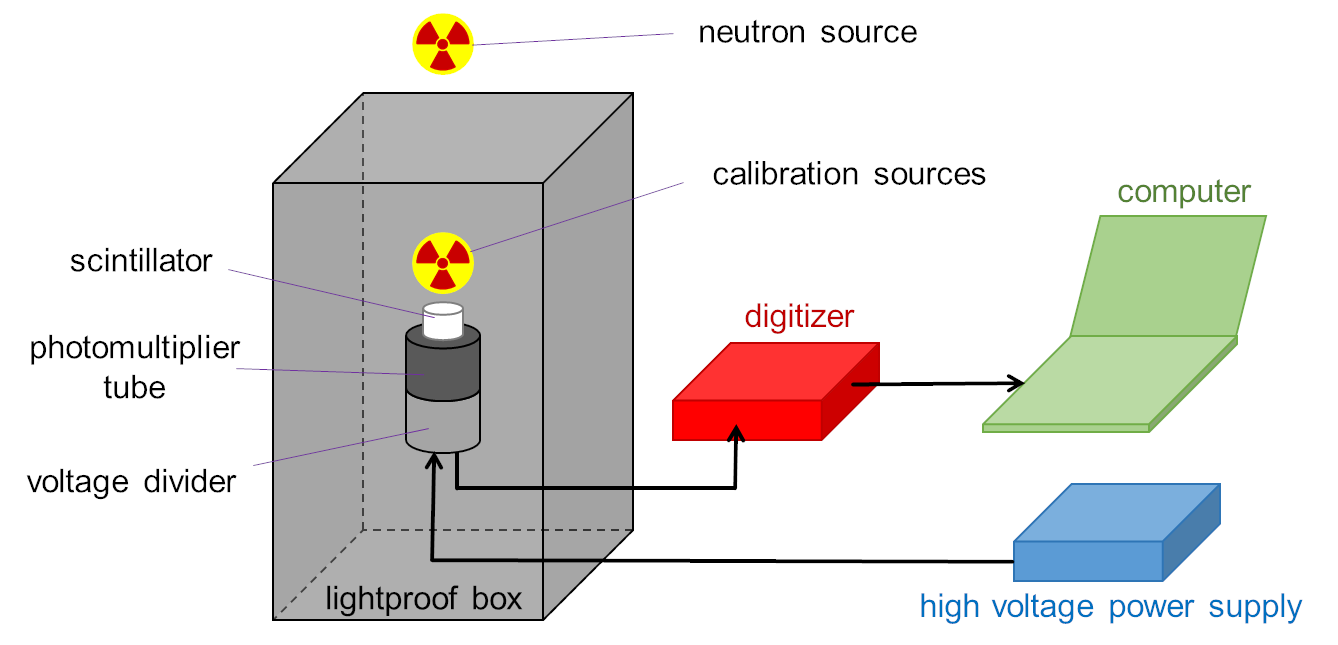}
    \caption{Experimental setup for \mbox{neutron-gamma} discrimination performance measurement.}
    \label{fig:experimental_setup_FOM}
\end{figure}

For the \mbox{neutron-gamma} pulse shape discrimination tests, we used samples attached to the photomultiplier in the same way as for photoelectron yield measurements, but the experimental setup was simpler (Fig. \ref{fig:experimental_setup_FOM}). The anode output of the PMT was connected directly to the CAEN DT5730 USB desktop digitizer \cite{CEANwebsite}. Raw waveforms were recorded in binary files using CoMPASS software. We used Constant Fraction Discrimination as the trigger method. The recorded data were analyzed using the Charge Comparison Method (CCM) implemented in a custom software written in Python programming language with NumPy, SciPy, and PyGAD libraries. Additionally we used the MatPlotLib library for the visualization of the results.
\par
All gamma sources mentioned at the beginning of this section were used for energy calibration, and the PuBe neutron source was used to measure the response of scintillators to both gamma radiation and fast neutrons.
\par

\subsection{Light pulse shape measurement setup} \label{sec:experimental_pulses}

\begin{figure}[!htb]
    \centering
    \includegraphics[width=0.9\linewidth]{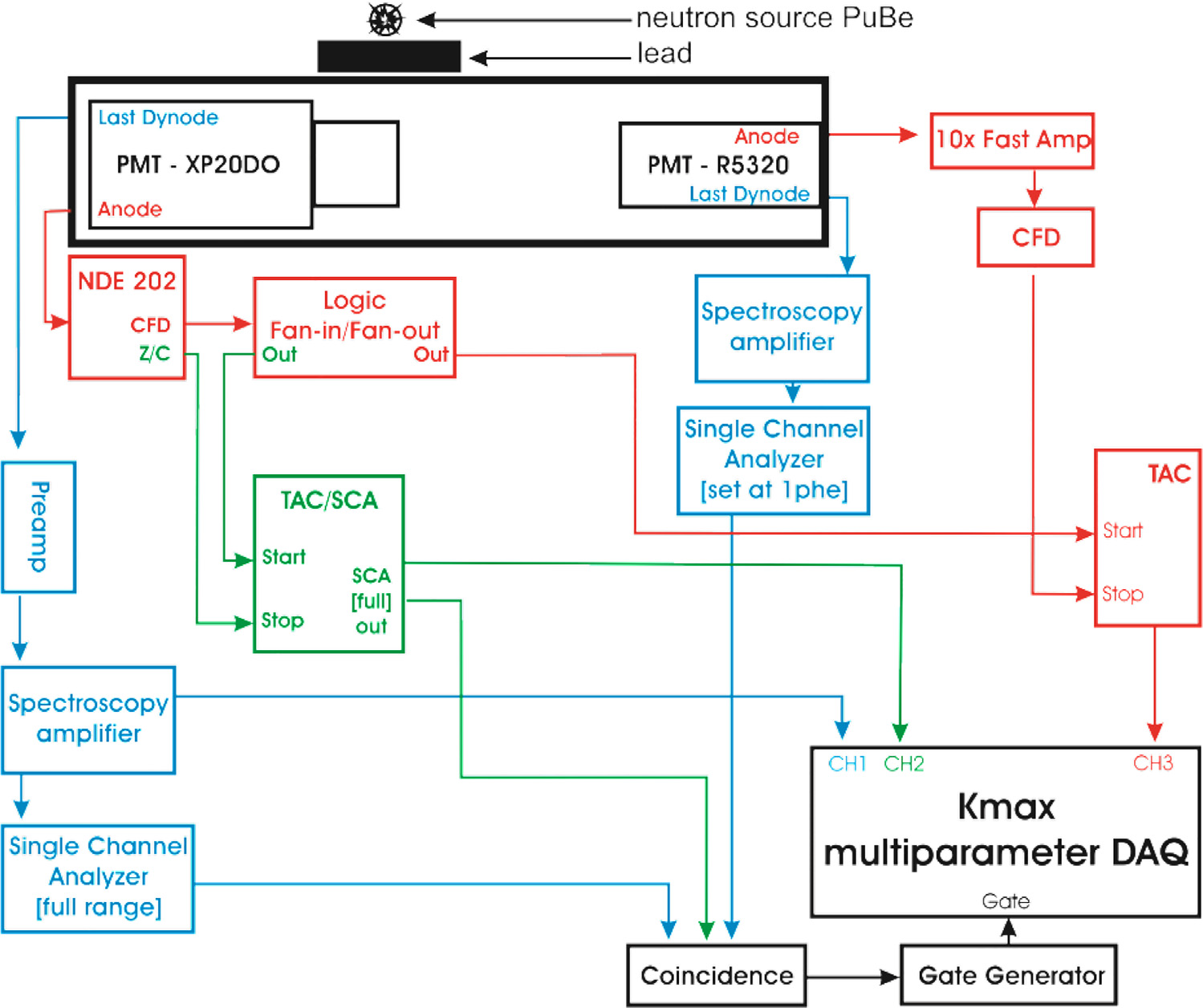}
    \caption{Experimental setup for light pulse shape measurement.}
    \label{fig:experimental_setup_pulses}
\end{figure}

We used the modified \mbox{Bollinger-Thomas} single photon method \cite{BollingerThomas1961} to record light pulse shapes of two scintillators: 25~mm OGS and 1-inch \mbox{trans-stilbene}. The slow-fast experimental setup details are presented in Fig. \ref{fig:experimental_setup_pulses}. In these measurements only the sides of the cylindrical scintillator samples were covered with Teflon tape, leaving the top and the bottom open to emit photons. The sample was attached to a Photonis XP20D0 photomultiplier to detect light pulses. Another photomultiplier (Hamamatsu R5320) was placed at distance around 10 cm on the other side of the scintillator to detect single photons from scintillations.
All these parts of the setup were inside of a lightproof box.
The time difference between the signals from both PMTs was converted to amplitude in a TAC (\mbox{time-amplitude} converter) and recorded along with the amplitude of the pulse (which was used to measure the energy) and the \mbox{zero-crossing} signal (which was used for \mbox{neutron-gamma} discrimination).
\par
Once again, we used a PuBe neutron source to obtain pulse shapes for both gamma and fast neutron radiation. It was placed behind a 5~cm thick lead brick to reduce the amount of \mbox{gamma-rays}, which could produce Cherenkov photons in a PMT window. \tsup{137}Cs and \tsup{22}Na sources placed next to the scintillator with no additional shielding were used for energy calibration in this experimental setup.
\par

\section{Results} \label{sec:results}

\subsection{Photoelectron yield and pulse shape discrimination} \label{sec:results_PSD}

The number of photoelectrons per MeV of all tested OGS samples is presented in Table \ref{tab:yield_FOM_FOMnorm_vs_H}, along with other parameters. Each measurement was repeated at least three times for every sample, with the scintillators reattached to the photomultiplier each time. The values in the table are average of all measurements.
\par
\begin{figure}[!htb]
    \centering
    \includegraphics[width=0.7\linewidth]{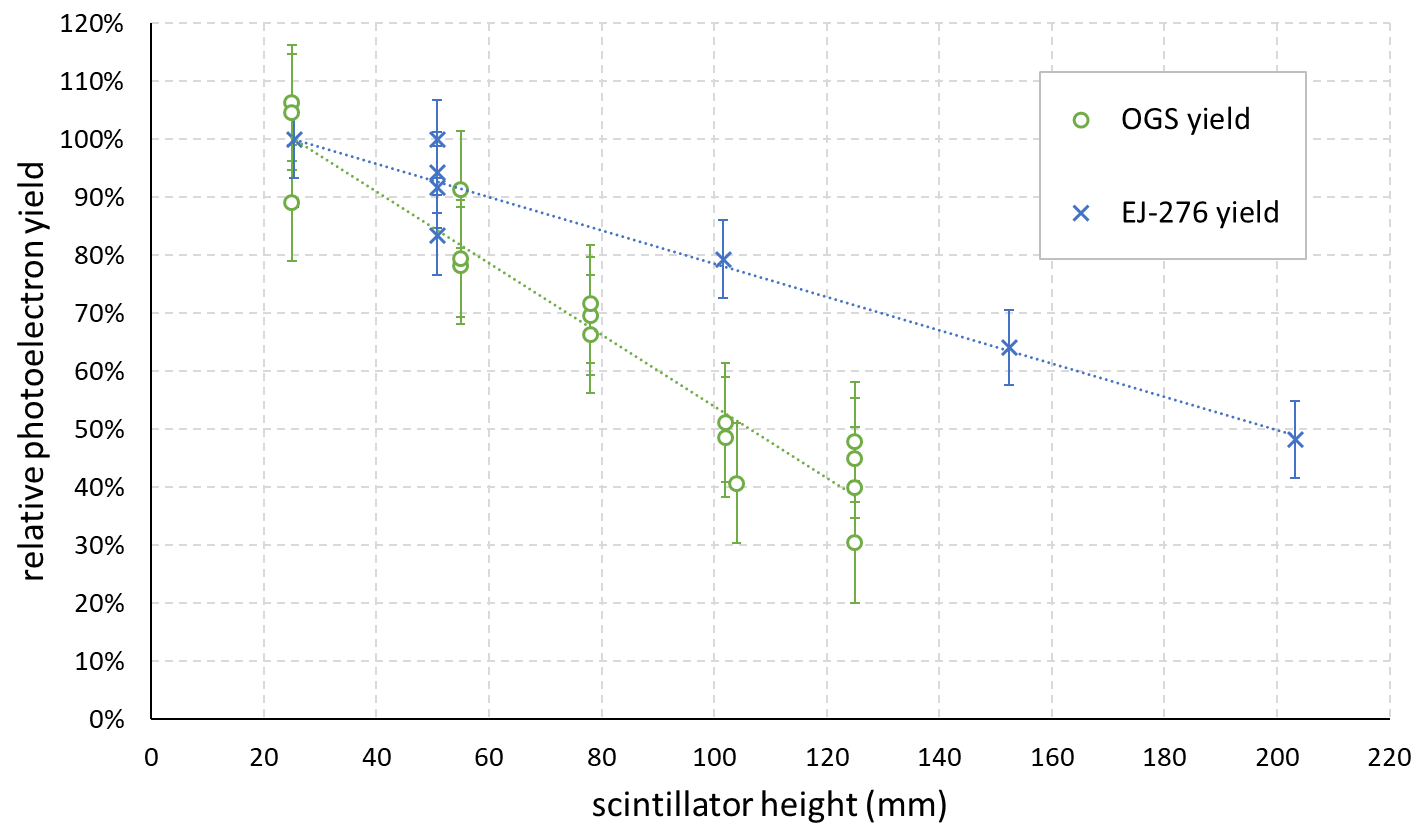}
    \caption{The decrease of photoelectron yield of OGS (green circles) compared to EJ-276 (blue crosses) with the scintillator height. EJ-276 data taken from \cite{GrodzickaKobylka2020JINST}. The yields were normalized to 100\% at 25.4~mm (1 inch). The uncertainties of OGS relative photoelectron yield represent 1$\sigma$ confidence level and were estimated from the variance of the results obtained for given sample and the variances of every measurement. One of the points of 102-mm OGS sample is shifted, because the sample was initially longer (104~mm) and mechanical damage after the first measurement resulted in shorter sample for the rest of the research.}
    \label{fig:yield_vs_H}
\end{figure}

Based on number of photoelectrons, we determined relative photoelectron yield, presented in Figure \ref{fig:yield_vs_H}. For comparison, we also included data from our previous study on the EJ276 plastic scintillator \cite{GrodzickaKobylka2020JINST}. In both cases, the yield decreases with increasing scintillator height; however, this effect is significantly more pronounced in OGS. We attribute this to stronger self-absorption of light in taller scintillators, which appears to be more pronounced in OGS than in EJ-276.
\par
Another difference is that EJ-276 plastics were stacked on the top of each other to measure height dependence and their faces were joined using silicone grease, while OGS samples were uniform with no joints. Despite this difference, the decrease of yield is linear in both cases, which ensures us that there was no significant loss of light on the surfaces of stacked EJ-276 plastics.
\par
It is worth mentioning that the relation between photoelectron yield and size of the scintillator is approximately linear. In case of absorption of light one may expect it to be exponential, but the absorption is not the only phenomenon affecting the results. The light is also produced in the scintillator itself and reflected from the walls of it, so the path of light through absorbing material depends on where and how the radiation interacts with the scintillator. This is related to the range of radiation in the scintillator material and the position of the source. In our case the \tsup{136}Cs source was always on the top of the scintillator, opposing the bottom coupled with the photomultiplier, resulting in linear relation.

\begin{table}[p]
\centering
\small
\begin{tabular}{ *{4}{c|}c }
    \hline\hline
height & photoelectron yield & \multirow{2}{*}{N\tsub{phe}@300keVee} & \multirow{2}{*}{FOM} & \multirow{2}{*}{$FOM/\sqrt{N_{phe}}$} \\
{[mm]} & {[phe/MeV]} &  &  & \\
    \hline
25 & 4310~\textpm~420 & 1290~\textpm~130 & 2.37~\textpm~0.07 & 0.066~\textpm~0.005 \\
55 & 3570~\textpm~320 & 1072~\textpm~97 & 2.11~\textpm~0.06 & 0.064~\textpm~0.004 \\
78 & 2980~\textpm~140 & 894~\textpm~41 & 2.01~\textpm~0.06 & 0.067~\textpm~0.002 \\
102 & 2020~\textpm~240 & 606~\textpm~73 & 1.78~\textpm~0.05 & 0.072~\textpm~0.006 \\
125 & 1760~\textpm~330 & 530~\textpm~100 & 1.70~\textpm~0.05 & 0.074~\textpm~0.010 \\
    \hline\hline
\end{tabular}
\caption{FOM and normalized FOM values at 300 keVee for OGS of different heights. In each case FOM values were obtained for short gate equal 66 ns and long gate equal 350 ns. FOM uncertainty was estimated at 3\%. Photoelectron yield uncertainty was estimated at 1$\sigma$ confidence level from Poisson distribution of photoelectron number.}
\label{tab:yield_FOM_FOMnorm_vs_H}
\end{table}

\begin{figure}[!htb]
    \centering
    \includegraphics[width=0.7\linewidth]{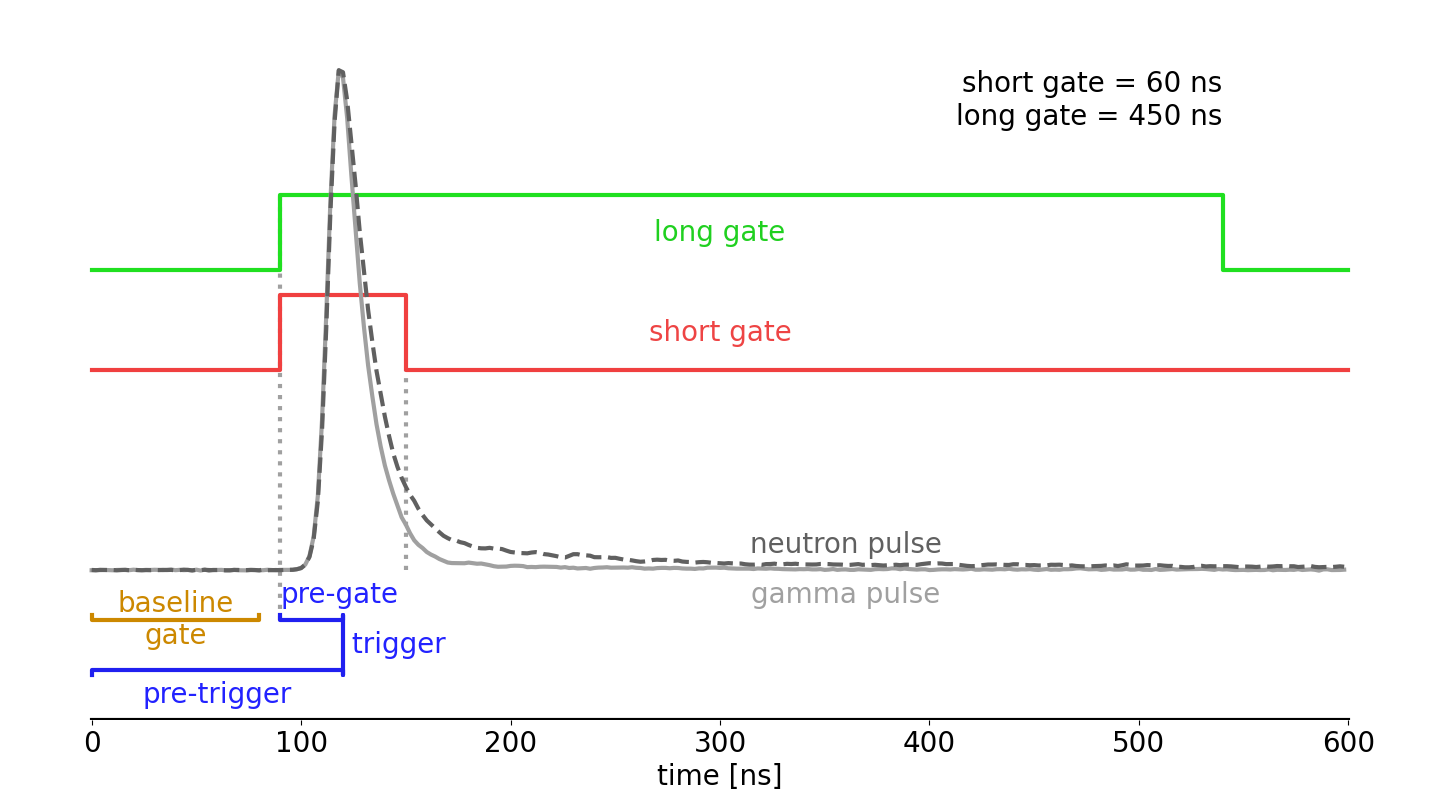}
    \caption{Example of gamma and neutron induced pulses recorded by digital analyzer with gates visualized. Long and short gate are the parameters of Charge Comparison Method. The trigger is dependent on the chosen triggering method and pre-gate is a parameter setting the beginning of the gates relative to it. Similarly, pre-trigger sets the beginning of recorded waveform. Baseline is calculated as an average of the data taken from baseline gate.}
    \label{fig:pulses}
\end{figure}

\begin{figure}[!htb]
    \centering
    \includegraphics[width=0.6\linewidth]{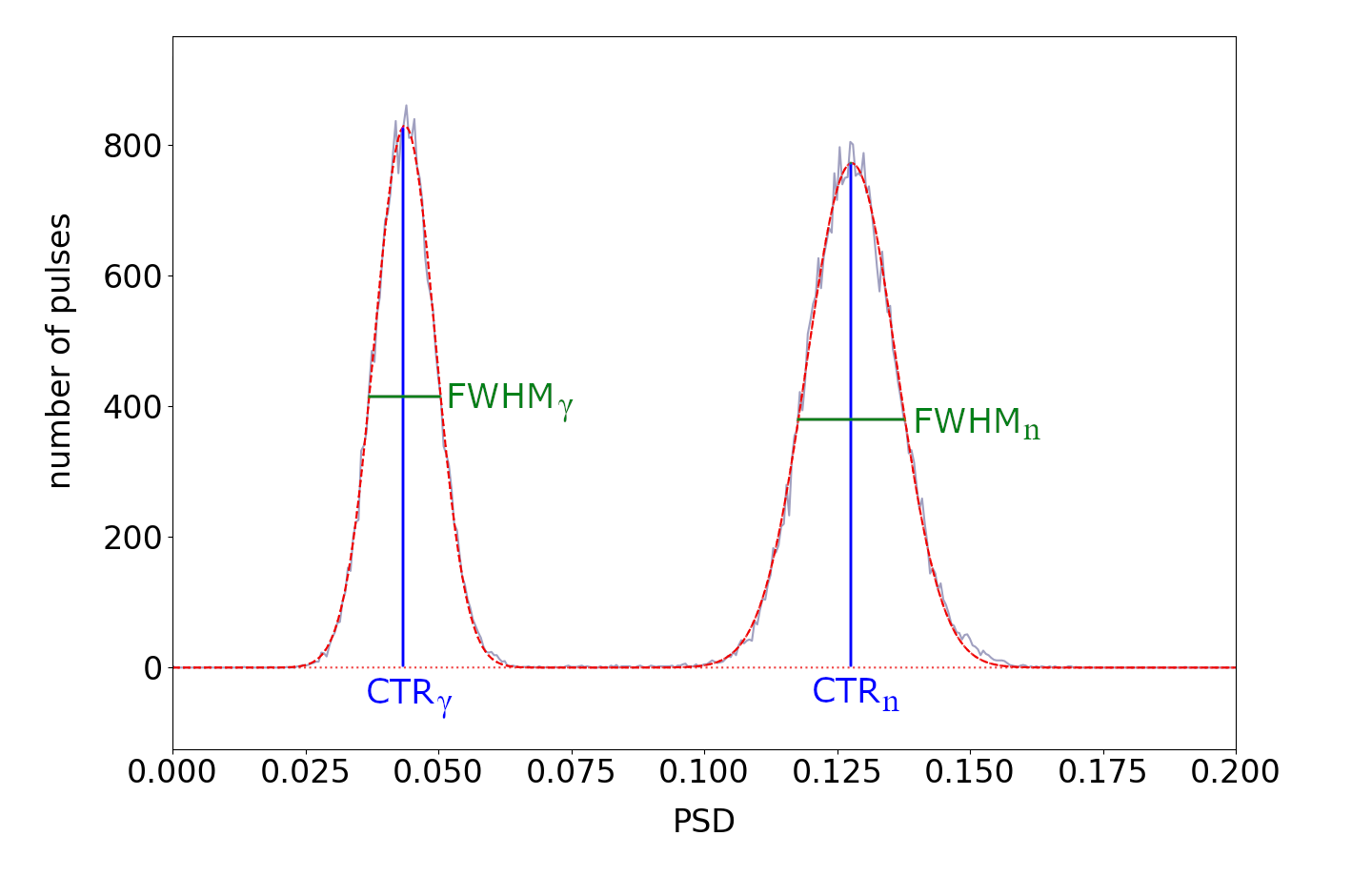}
    \caption{Example of PSD parameter histogram. CTR is a centroid of a peak and FWHM is a full width at half maximum of a peak. In case of the scintillators used in our research peak on the right side represents pulses induced by fast neutrons ($n$) and left peak pulses induced by gamma radiation ($\gamma$), because neutron pulses are longer, resulting in higher values of PSD parameter.}
    \label{fig:PSDhist}
\end{figure}

We also measured the Pulse Shape Discrimination (PSD) performance of all the OGS samples and compared it to \mbox{trans-stilbene} performance.
In order to quantify it, first we used Charge Comparison Method to calculate PSD parameter of each pulse, which was defined as a ratio of difference between integrals of a long gate ($Q_{long}$) and a short gate ($Q_{short}$) divided by the integral of a long gate (see Fig. \ref{fig:pulses}):
\begin{equation}
    PSD = \frac{Q_{long} - Q_{short}}{Q_{long}}
\end{equation}
Then we calculated the Figure of Merit (FOM) as follows:
\begin{equation}
    FOM = \frac{|CTR_{n}-CTR_{\gamma}|}{FWHM_{n}+FWHM_{\gamma}}
\end{equation}
where FWHM and CTR stand for the full width at half maximum and the centroid of the peaks, respectively, corresponding to neutron ($n$) and gamma~ray ($\gamma$) detection as projected onto the PSD parameter axis, see Fig. \ref{fig:PSDhist}. The higher the FOM, the better PSD performance.
\par
Energy calibration was based on an assumption that the energy $E$ is a function of the channel number $ch$, which is proportional to the long gate integral $Q_{long}$. This proved to be sufficient even when long gate covered only a part of a pulse. We used quadratic calibration function to address the non-proportionality of the scintillators.
\begin{equation}
    ch \sim Q_{long}
\end{equation}
\begin{equation}
    E = f(ch)
\end{equation}

There are several parameters in Charge Comparison Method that have to be tuned to obtain the best (maximum) FOM value. The most important parameters are the short and long gates lengths (Fig. \ref{fig:pulses}). As our experience shows, the moment the gates start should also not be neglected (it is determined by \textit{\mbox{pre-gate}} setting in CAEN digitizers, but we used another approach in \mbox{off-line} analysis of recorded waveforms). The triggering method and the way the baseline is calculated can also impact the results, so they should be chosen carefully (see \cite{Adamowski2025pulses} for more details).
\par
\begin{figure}[!htb]
    \centering
    \includegraphics[width=0.48\linewidth]{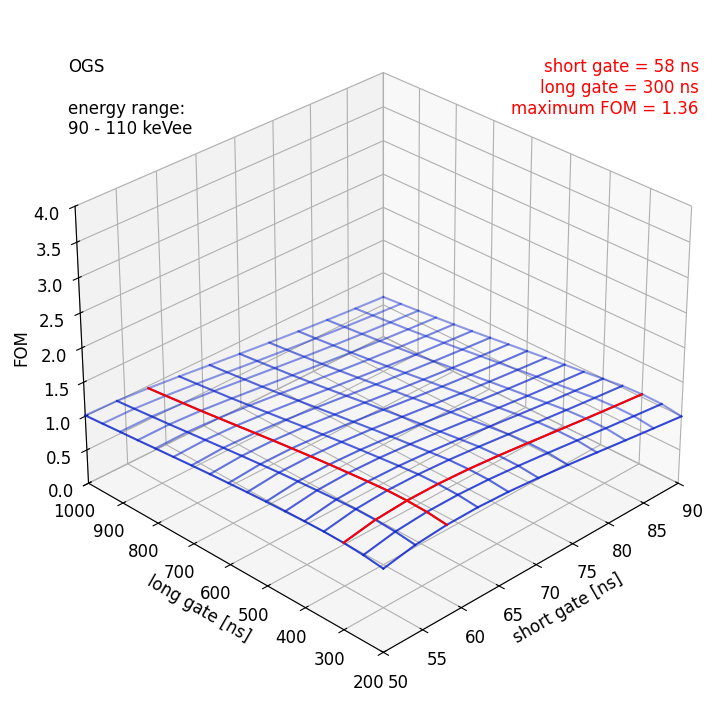}
    \includegraphics[width=0.48\linewidth]{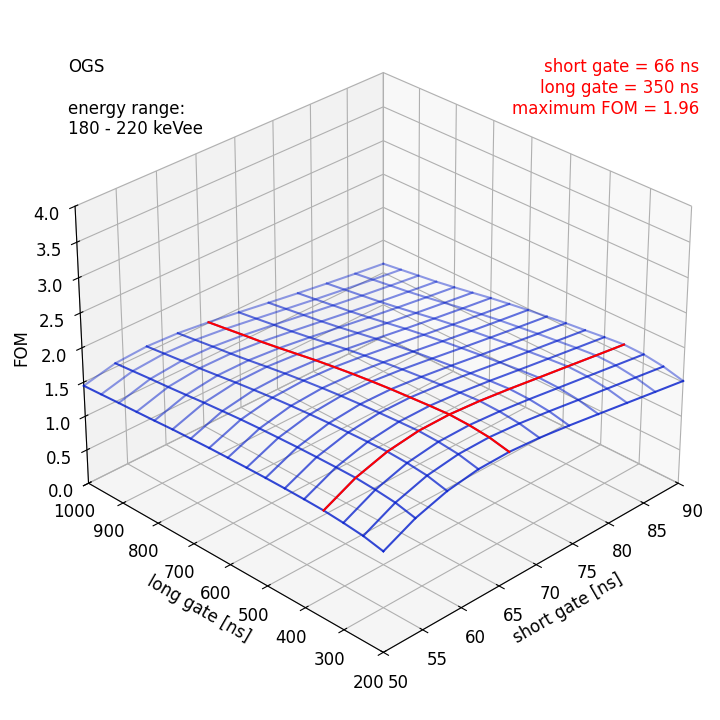}
    \includegraphics[width=0.48\linewidth]{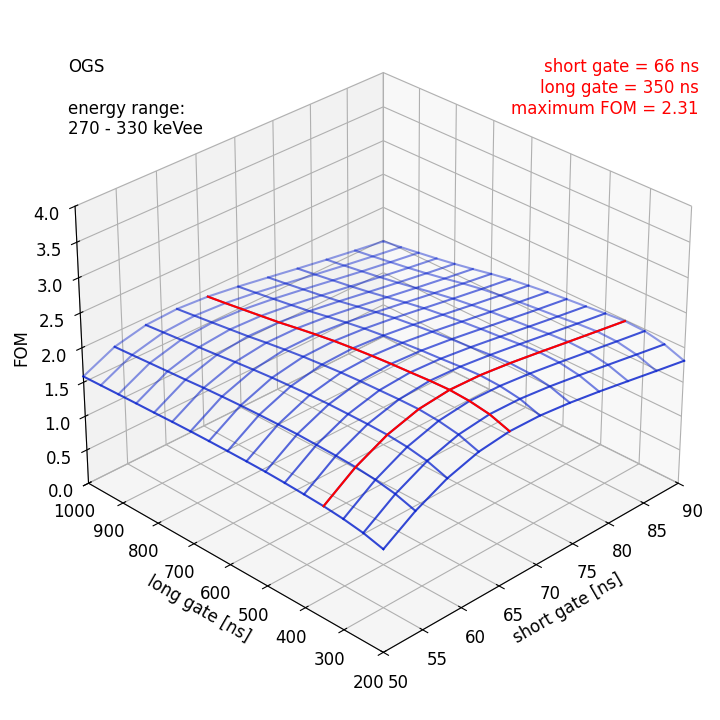}
    \includegraphics[width=0.48\linewidth]{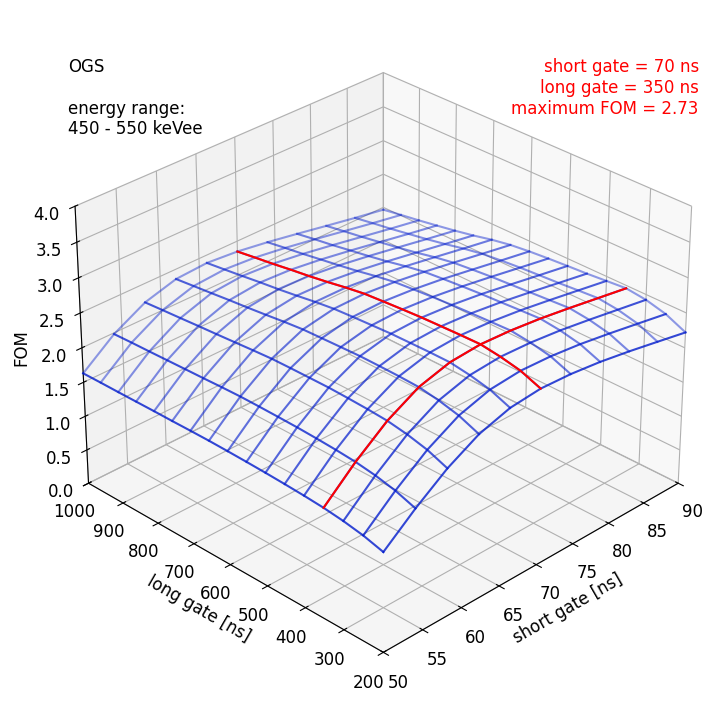}
    \includegraphics[width=0.48\linewidth]{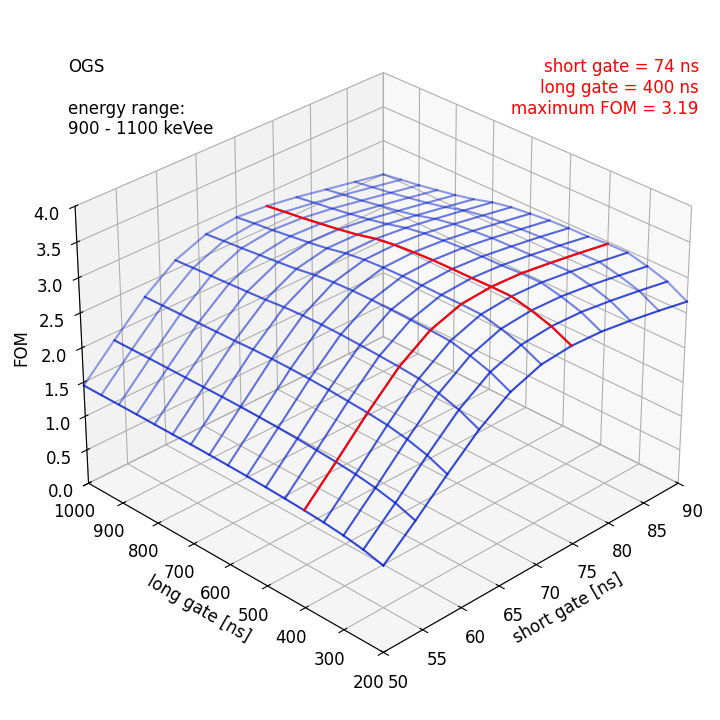}
    \includegraphics[width=0.48\linewidth]{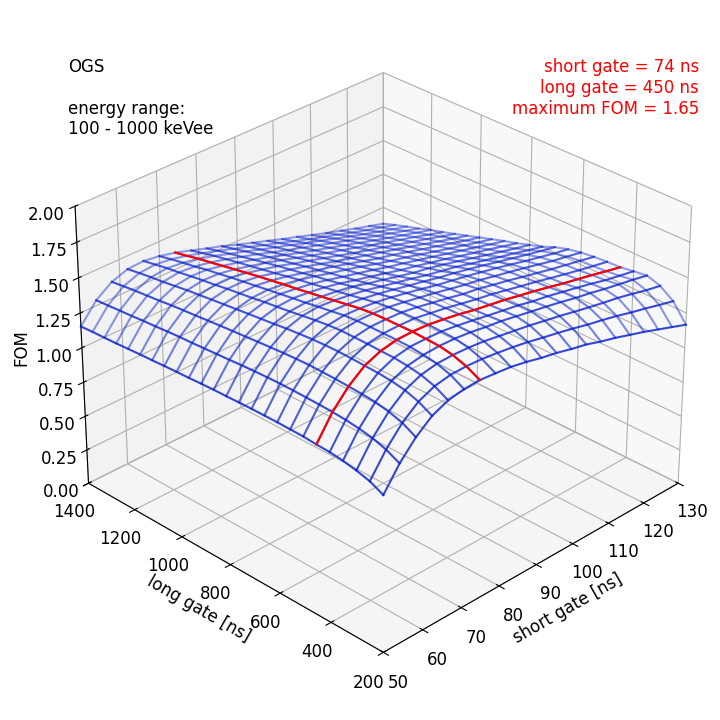}
    \caption{Example of FOM dependence on short and long gates for variety of energies in 25~mm organic glass scintillator. Best (maximum) FOM gates data are marked red.}
    \label{fig:FOM_3D}
\end{figure}

Manually tuning all these parameters is \mbox{time-consuming} and prone to mistakes, so we took advantage of the possibility of recording pulses in the form of digital waveforms and analyzing them \mbox{off-line}. We used custom software written in Python to obtain maps of FOM values for many different pairs of short and long gates and across many energy ranges (see Fig. \ref{fig:FOM_3D}). Short gates were scanned in 4~ns steps, and long gates in 50~ns steps. Both gates started 80~ns after the beginning of the waveform, i.e. approximately 25~ns before pulse maximum. The baseline of each pulse was estimated as average value of data in 80~ns long baseline gate, starting from the beginning of the waveform.
\par
We used {E}~\textpm~{0.1}\texttimes{E} pattern to define the energy range rather than a fixed range width because the number of pulses decreases with energy. Choosing a fixed width would result in fewer pulses at higher energies, leading to greater statistical noises, which could affect reliability of the results.
\par
As can be seen in Fig. \ref{fig:FOM_3D}, for each energy range, there is an optimal pair of short and long gate for which FOM value reaches its maximum (i.e. the PSD performance is the best). By averaging results from a series of measurements, we calculated best FOM values across many energy ranges for all OGS samples as well as for \mbox{trans-stilbene} crystal. The results are presented in Table \ref{tab:FOM_vs_E_vs_H}. We estimated uncertainty of FOM at 3\%, which is slightly higher than the highest FOM standard deviation we observed. If, in consecutive measurements of the same sample, optimal gates differed (usually due to statistical noises), we averaged gates as well. The averaged short gates were rounded to 2~ns, as this is the time resolution of gates in the CAEN digitizer we used. The averaged long gates were rounded to 10~ns, as the relatively high 50~ns step size made it impossible to achieve better resolution.
\par
\begin{figure}[!htb]
    \centering
    \includegraphics[width=0.49\linewidth]{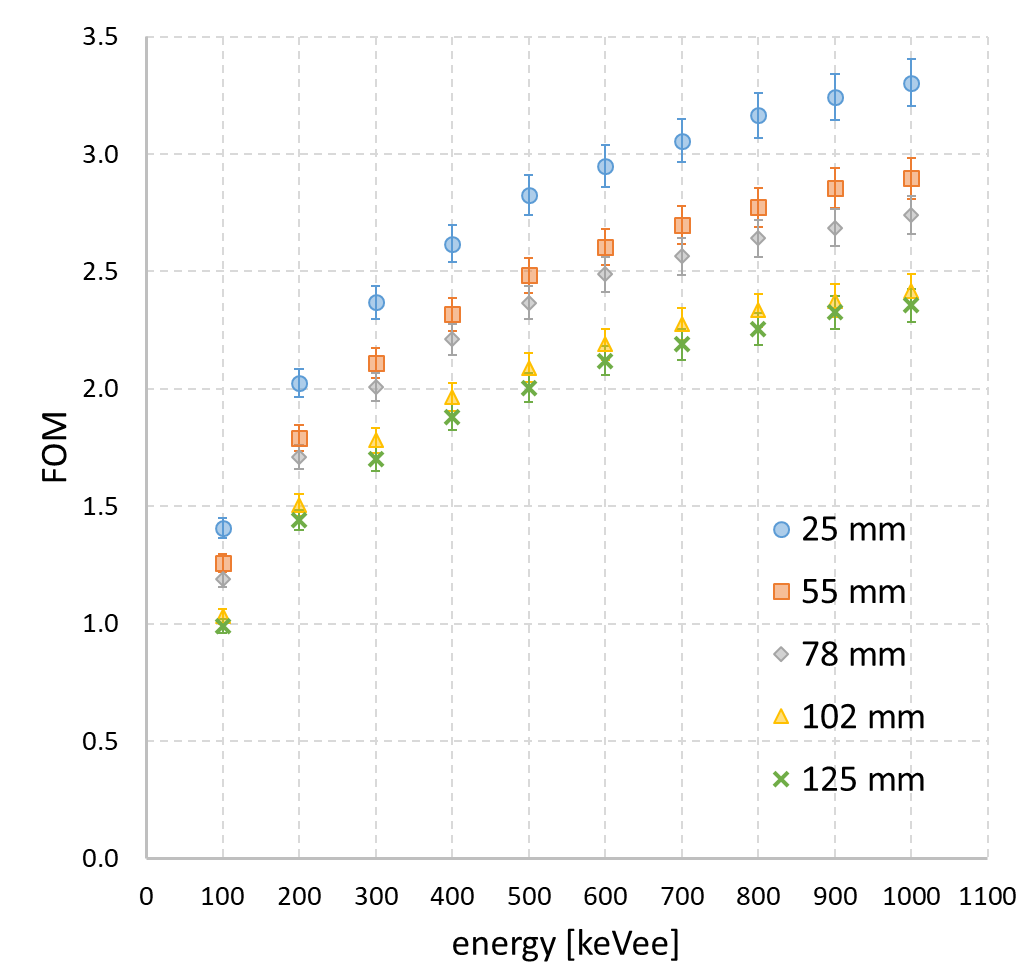}
    \includegraphics[width=0.49\linewidth]{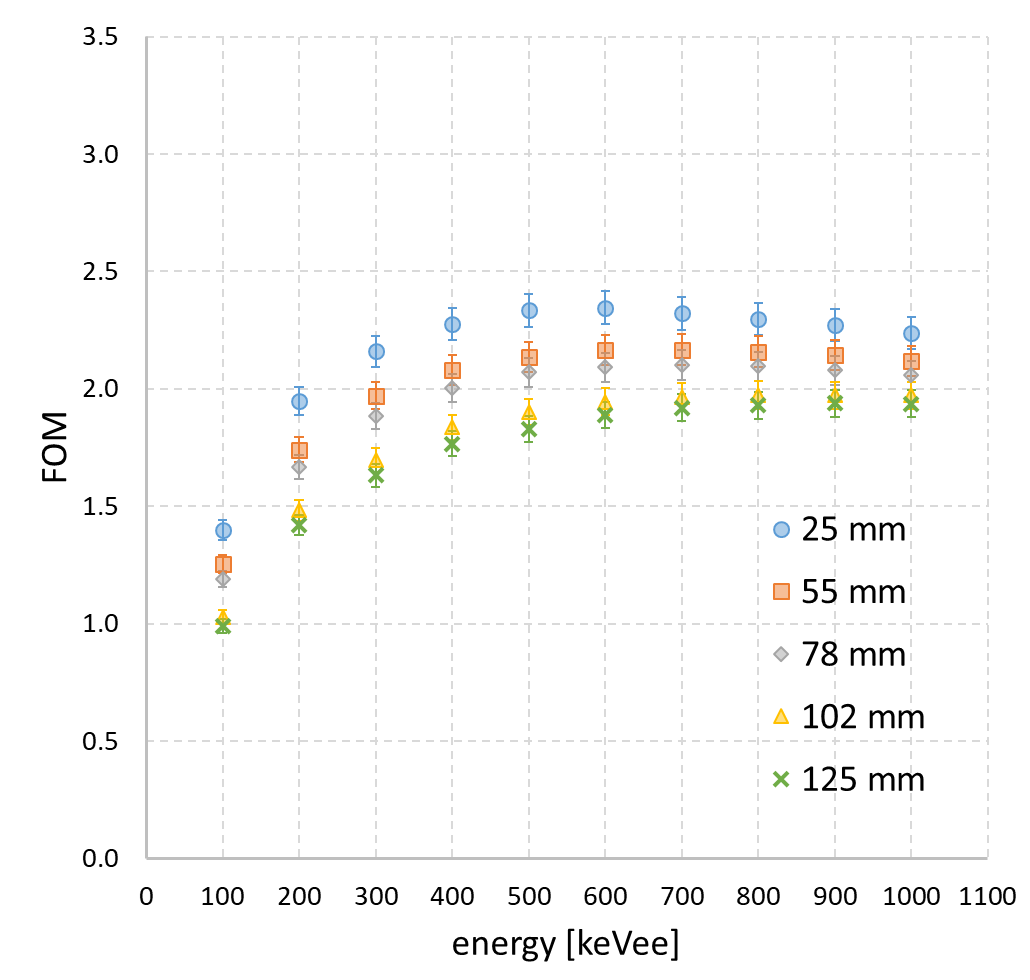}
    \caption{Average best FOM (left) and FOM at fixed gates short = 58~ns, long = 300~ns (right) of OGS samples at different energy ranges.
    The best gates at the lowest energy (100~keVee) were chosen as the fixed gates, because FOM values of all OGS samples are the lowest at this energy.
    With these gates, PSD capability at higher energies is not impacted significantly, since FOM values above 2 mean almost complete separation of neutron and gamma radiation pulses. Choosing other gates would result in worsening PSD capability at 100~keVee, which is significant, as FOM values at this energy are already below 1.5. The uncertainty of FOM values was estimated at 3\% from standard deviation of all measurements.}
    \label{fig:FOM_vs_E}
\end{figure}

It must be stressed, that optimal gates change with the energy. We observed it in both OGS and \mbox{trans-stilbene}. Presumably this is also true for other scintillators capable of neutron-gamma discrimination. This means that in practical applications the gates need to be tuned depending on the energy range one wants to measure. It is quite simple when the energy range is narrow, like in our case. When energy range is wide, we advise choosing gates corresponding to the lowest energy. The reason for this choice is that FOM values are lower at lower energies, sometimes going near the threshold at which neutron and gamma ray pulses can no longer be distinguished, causing a critical failure.
Choosing different gates could result in such failure at this low energy, while using sub-optimal gates at higher energies does not worsen PSD capability significantly, since FOM values are much higher and far from being critical. Therefore, choosing gates corresponding to the lowest energy would minimize the risk of failure at low energies, while retaining good neutron-gamma discrimination capability at high energies (see Fig. \ref{fig:FOM_vs_E} right).

\par
\begin{figure}[!htb]
    \centering
    \includegraphics[width=0.5\linewidth]{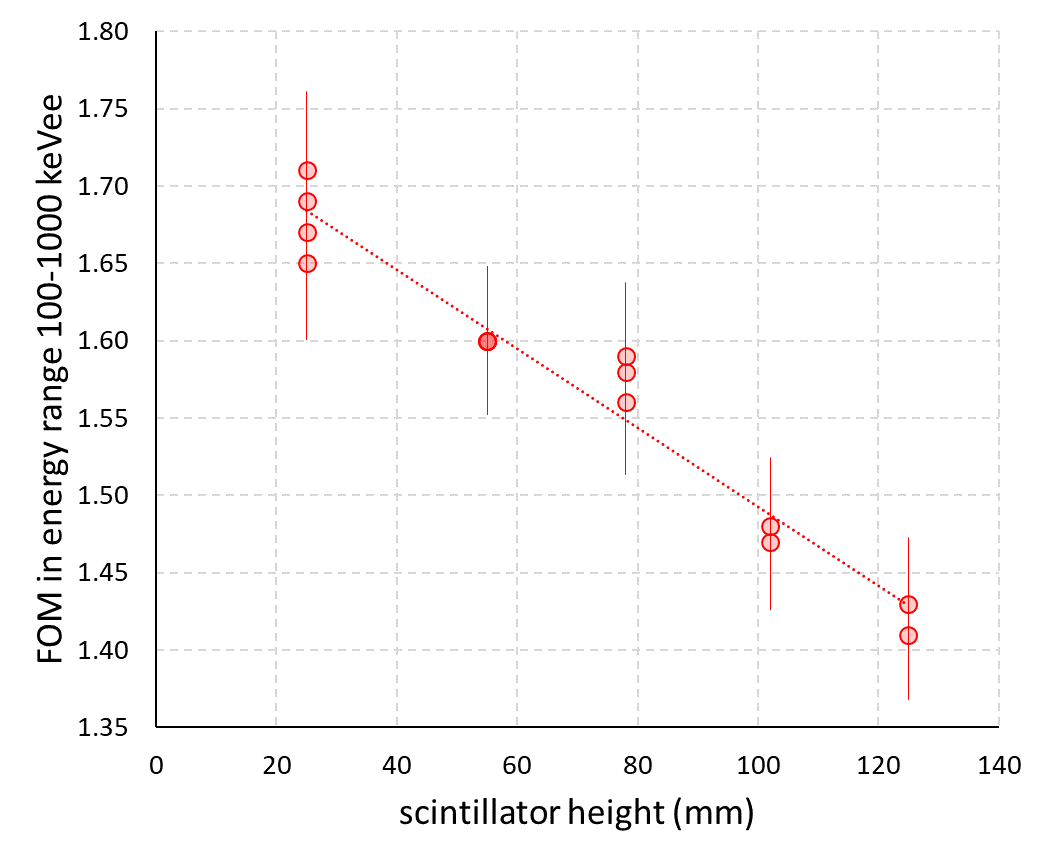}
    \caption{OGS FOM in wide energy range 100-1000~keVee decreases with scintillator height. Several measurements were made with every sample. An apparent single point on the graph for 55~mm sample is a result of three independent measurements, all with the same FOM value 1.6. The uncertainty of FOM values was estimated at 3\% from standard deviation of all measurements.}
    \label{fig:FOM_vs_H}
\end{figure}

As it can be seen in Figure \ref{fig:FOM_vs_H}, Figure \ref{fig:FOM_vs_E} and Table \ref{tab:FOM_vs_E_vs_H}, FOM values decrease with the height of the scintillator in a manner similar to the decrease in the number of photoelectrons. This supports the idea that light \mbox{self-absorption} in organic scintillators reduces their \mbox{neutron-gamma} discrimination capability and imposes a limit on the maximum useful detector size.
\par
Table \ref{tab:yield_FOM_FOMnorm_vs_H} summarizes the photoelectron yield, FOM values at an energy of 300~keVee, and FOM values normalized to the number of photoelectrons ($FOM/\sqrt{N_{phe}}$) at the same energy. Despite changes in FOM, the normalized FOM values are similar regardless of the size of the scintillator, within the range of uncertainties. Therefore, normalized FOM at given energy can be considered an intrinsic property of the scintillator.

\subsection{Light pulse shapes} \label{sec:results_pulses}

\begin{figure}[!htb]
    \centering
    \includegraphics[width=0.49\linewidth]{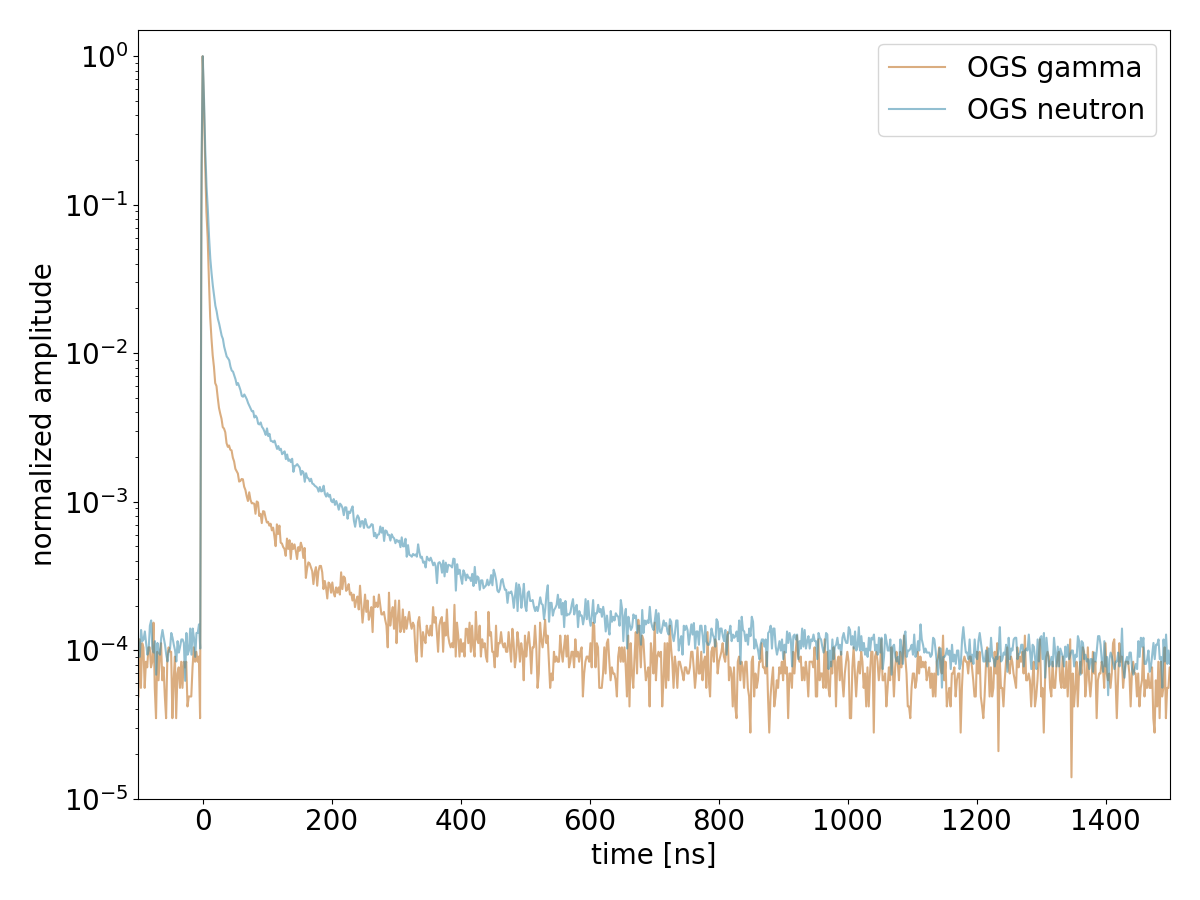}
    \includegraphics[width=0.49\linewidth]{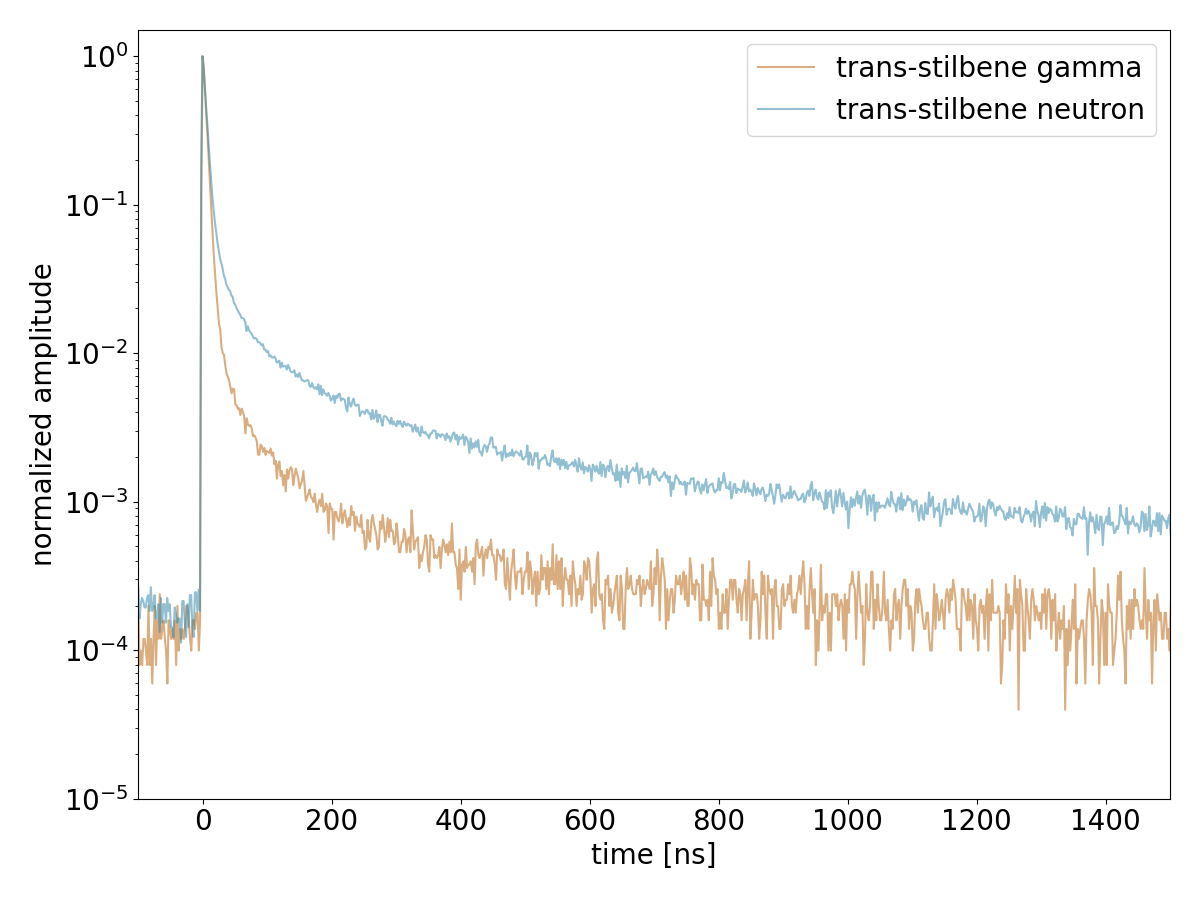}
    \caption{The pulse shapes of gamma and neutron induced pulses in OGS (left) and \mbox{trans-stilbene} (right) for a wide energy range from 350~keVee to 1250~keVee. Data shown with sampling at time resolution 2000~ns per 1024 channels.}
    \label{fig:pulses_gamma_neutron}
\end{figure}

Data recorded with \mbox{Bollinger-Thomas} setup were used to investigate light pulse shapes of 25~mm OGS and 25.4~mm (1 inch) \mbox{trans-stilbene}. We were able to separate neutron and gamma events using zero crossing information. Energy calibration done with \tsup{137}Cs and \tsup{22}Na gamma sources let us also select events from a specific energy range. Unfortunately, we could get reliable data in limited energy ranges and only relatively small range of energies from 350~keVee to 1250~keVee was common for both scintillators. Nonetheless, in this range there is a clear difference between gamma and fast neutron signals in both of them (Fig. \ref{fig:pulses_gamma_neutron}).
\par

We estimated decay parameters by fitting function $f$ to the data:
\begin{equation}
    f(t) = C + A_1 e^{-t/\tau_1} + A_2 e^{-t/\tau_2} + A_3 e^{-t/\tau_3}
\end{equation}
where $t$ is time starting from the maximum of the pulse, $A_1$, $A_2$ and $A_3$ are the amplitudes of the exponential functions, $\tau_1$, $\tau_2$ and $\tau_3$ are the scintillation decay time constants, and $C$ is the baseline offset. We also calculated the intensity $I$ of each component:
\begin{equation}
    I_k = \frac{A_k \times \tau_k}{\sum_i A_i \times \tau_i}
\end{equation}
\par
To estimate $A$, $\tau$, and $C$ parameters we used Monte Carlo searching method in a form of Genetic Algorithm (GA) \cite{Katoch2021} implemented in Python PyGAD library \cite{PyGADwebsite}. We used the reciprocal of mean squared residuals as a fitting function of GA, where residual was defined as the difference between the observed value $y$ and predicted value $f$, divided by uncertainty of $y$. In other words, it was GA implementation of weighted least squares parameter estimation. The uncertainty of $y$ was estimated at $\sqrt{y}$, since each data point recorded in \mbox{Bollinger-Thomas} setup follows Poisson distribution.
\par
The advantage of GA is that it is capable of avoiding local optima, because initial values of the parameters are randomized (within defined ranges) and the process of approaching optimum also involves random steps.
This is an improvement over iterative methods like deterministic Levenberg–Marquardt algorithm (which is implemented e.g. in Origin software and in Python SciPy library \textit{curve\_fit} function), that rely heavily on manually set initial values, tend to find local optima and sometimes loop indefinitely.
Using GA we could make fitting procedure more automatic and minimize the impact of human factors.
\par
\begin{figure}[!htb]
    \centering
    \includegraphics[width=0.98\linewidth]{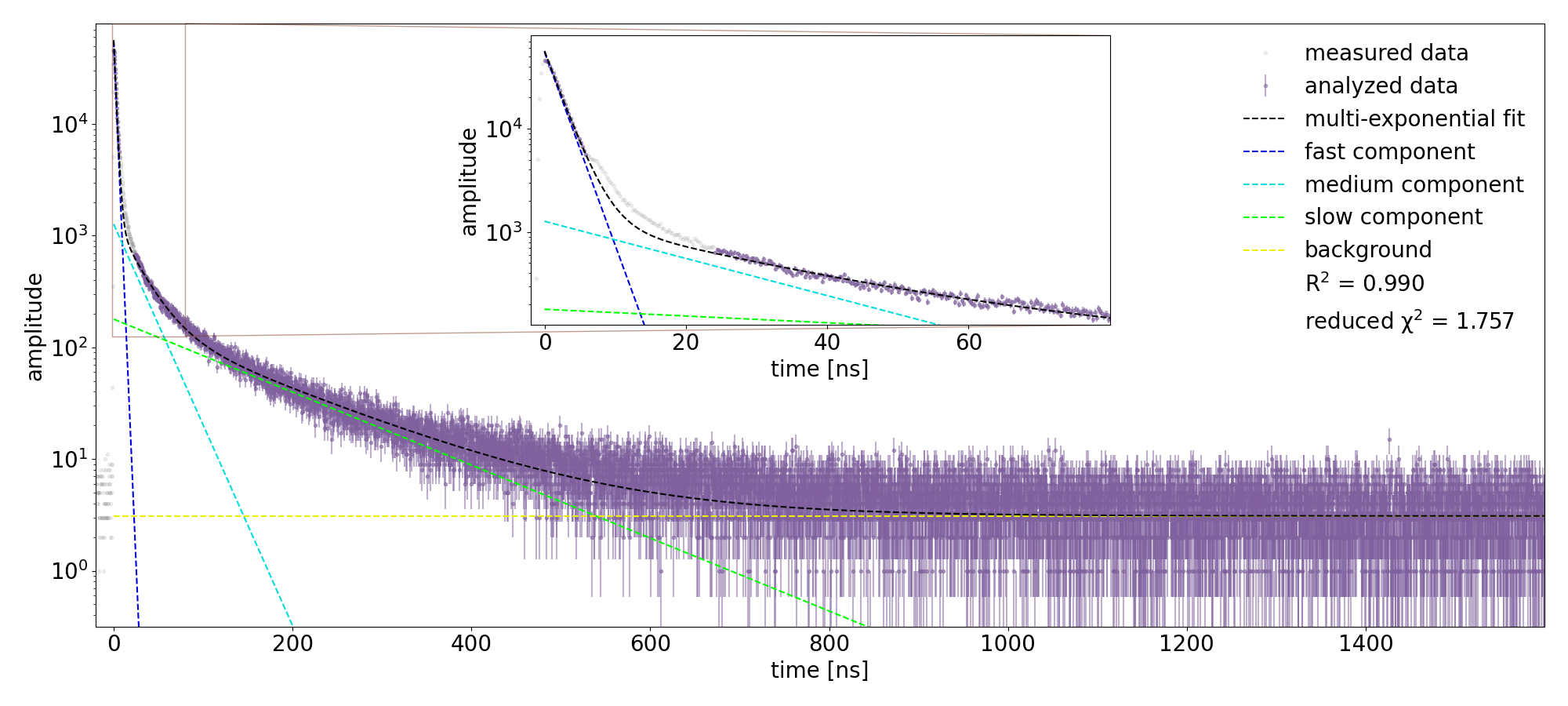}
    \includegraphics[width=0.98\linewidth]{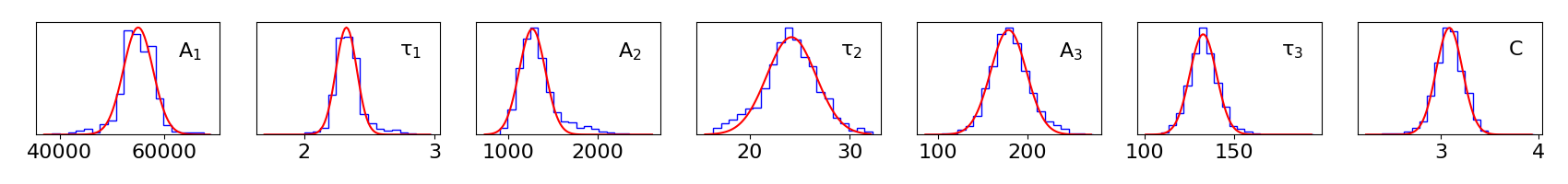}
    \caption{Example of the pulse shape (top) and the distributions of parameters (bottom), in case of neutron induced pulses in OGS for a wide energy range from 350~keVee to 1250~keVee.
    The bottom row figures show each parameter distribution (blue lines) and the result of fitting Gaussian function to this distribution (red lines) to obtain final parameter value and uncertainty.
    Top figure shows the final result of multi-exponential function fitting, with fast, medium, and slow component depicted, as well as the estimated background.
    Data were sampled at time resolution 2000~ns per 8192 channels.
    Data used to estimate the parameters (purple dots) are presented along with their uncertainties.
    $R^2$ and reduced $\chi^2$ values were calculated to confirm that the results are reliable.
    }
    \label{fig:pulse_OGS_neutron}
\end{figure}

On the other hand, using randomization means that the end results of GA algorithm are also affected by a random noise or sometimes simply incorrect.
Therefore, we ran this algorithm at least several hundred times for each pulse to obtain a distribution of values for every parameter (see blue lines in the bottom of Figure \ref{fig:pulse_OGS_neutron}). These distributions were corrected by removing obviously false results (e.g. with parameter values out of the defined range).
Additionally, we wanted to take into account the fact that the choice of data also affects the parameters, so only randomly taken 20\% of data were analyzed for each 10 runs. Based on these distributions we estimated parameters and their uncertainties by fitting Gaussian curves (the same Figure, red lines).
The results are in the Table \ref{tab:pulse_components}.
\par
As we have shown in Section \ref{sec:results_PSD}, optimal short and long gates change with energy. To investigate the reason for this, we estimated decay parameters of pulses in narrow energy ranges from 200~keVee to 1000~keVee, using the same GA approach as for wide energy range.
We found no significant change in decay times across all energies in both scintillators, neither for neutron- nor gamma-induced pulses.

\begin{table}[p]
\centering
\small
\begin{tabular}{ c|c|c|*{5}{c} }
    \hline\hline
energy &  & stilbene & OGS & OGS & OGS & OGS & OGS \\
{[keVee]} &  & 1~inch & 25~mm & 55~mm & 78~mm & 102~mm & 125~mm \\
    \hline\hline
 & short gate & 62 ns & 58 ns & 58 ns & 58 ns & 58 ns & 58 ns \\
100~\textpm~10 & long gate & 400 ns & 270 ns & 280 ns & 280 ns & 300 ns & 300 ns \\
 & FOM & 2.43~\textpm~0.07 & 1.41~\textpm~0.04 & 1.26~\textpm~0.04 & 1.19~\textpm~0.04 & 1.03~\textpm~0.03 & 0.99~\textpm~0.03 \\
    \hline
 & short gate & 70 ns & 64 ns & 62 ns & 66 ns & 62 ns & 62 ns \\
200~\textpm~20 & long gate & 550 ns & 320 ns & 320 ns & 320 ns & 320 ns & 350 ns \\
 & FOM & 3.53~\textpm~0.11 & 2.02~\textpm~0.06 & 1.79~\textpm~0.05 & 1.71~\textpm~0.05 & 1.51~\textpm~0.05 & 1.44~\textpm~0.04 \\
    \hline
 & short gate & 70 ns & 66 ns & 66 ns & 66 ns & 66 ns & 64 ns \\
300~\textpm~30 & long gate & 700 ns & 330 ns & 330 ns & 330 ns & 350 ns & 350 ns \\
 & FOM & 4.14~\textpm~0.12 & 2.37~\textpm~0.07 & 2.11~\textpm~0.06 & 2.01~\textpm~0.06 & 1.78~\textpm~0.05 & 1.70~\textpm~0.05 \\
    \hline
 & short gate & 74 ns & 68 ns & 66 ns & 66 ns & 66 ns & 66 ns \\
400~\textpm~40 & long gate & 700 ns & 350 ns & 350 ns & 350 ns & 350 ns & 350 ns \\
 & FOM & 4.60~\textpm~0.14 & 2.62~\textpm~0.08 & 2.32~\textpm~0.07 & 2.21~\textpm~0.07 & 1.97~\textpm~0.06 & 1.88~\textpm~0.06 \\
    \hline
 & short gate & 74 ns & 70 ns & 70 ns & 70 ns & 66 ns & 66 ns \\
500~\textpm~50 & long gate & 750 ns & 350 ns & 350 ns & 350 ns & 380 ns & 350 ns \\
 & FOM & 4.95~\textpm~0.15 & 2.83~\textpm~0.08 & 2.48~\textpm~0.07 & 2.37~\textpm~0.07 & 2.09~\textpm~0.06 & 2.01~\textpm~0.06 \\
    \hline
 & short gate & 78 ns & 70 ns & 70 ns & 70 ns & 70 ns & 68 ns \\
600~\textpm~60 & long gate & 700 ns & 380 ns & 350 ns & 350 ns & 350 ns & 350 ns \\
 & FOM & 5.17~\textpm~0.16 & 2.95~\textpm~0.09 & 2.60~\textpm~0.08 & 2.49~\textpm~0.07 & 2.19~\textpm~0.07 & 2.12~\textpm~0.06 \\
    \hline
 & short gate & 78 ns & 72 ns & 70 ns & 70 ns & 70 ns & 70 ns \\
700~\textpm~70 & long gate & 1000 ns & 360 ns & 370 ns & 350 ns & 350 ns & 350 ns \\
 & FOM & 5.40~\textpm~0.16 & 3.06~\textpm~0.09 & 2.70~\textpm~0.08 & 2.56~\textpm~0.08 & 2.28~\textpm~0.07 & 2.19~\textpm~0.07 \\
    \hline
 & short gate & 78 ns & 74 ns & 70 ns & 70 ns & 70 ns & 70 ns \\
800~\textpm~80 & long gate & 1000 ns & 360 ns & 370 ns & 350 ns & 350 ns & 350 ns \\
 & FOM & 5.60~\textpm~0.17 & 3.16~\textpm~0.09 & 2.77~\textpm~0.08 & 2.64~\textpm~0.08 & 2.34~\textpm~0.07 & 2.26~\textpm~0.07 \\
    \hline
 & short gate & 78 ns & 74 ns & 74 ns & 74 ns & 70 ns & 70 ns \\
900~\textpm~90 & long gate & 1100 ns & 370 ns & 370 ns & 350 ns & 350 ns & 350 ns \\
 & FOM & 5.78~\textpm~0.17 & 3.24~\textpm~0.10 & 2.85~\textpm~0.09 & 2.69~\textpm~0.08 & 2.38~\textpm~0.07 & 2.33~\textpm~0.07 \\
    \hline
 & short gate & 78 ns & 74 ns & 74 ns & 74 ns & 70 ns & 70 ns \\
1000~\textpm~100 & long gate & 1100 ns & 380 ns & 370 ns & 350 ns & 350 ns & 350 ns \\
 & FOM & 5.90~\textpm~0.18 & 3.30~\textpm~0.10 & 2.90~\textpm~0.09 & 2.74~\textpm~0.08 & 2.42~\textpm~0.07 & 2.36~\textpm~0.07 \\
    \hline\hline
 & short gate & 78 ns & 74 ns & 72 ns & 70 ns & 70 ns & 70 ns \\
100~--~1000 & long gate & 1300 ns & 420 ns & 400 ns & 400 ns & 380 ns & 350 ns \\
 & FOM & 2.25~\textpm~0.07 & 1.68~\textpm~0.05 & 1.60~\textpm~0.05 & 1.58~\textpm~0.05 & 1.48~\textpm~0.04 & 1.42~\textpm~0.04 \\
    \hline\hline
\end{tabular}
\caption{Maximum FOM values at given energy for 1 inch high \mbox{trans-stilbene} crystal and OGS samples of different heights. The uncertainty of FOM was estimated at 3\% from FOM standard deviation. The uncertainties of the gates can be estimated as the step size: 4 ns for short gate, 50 ns for long gate.}
\label{tab:FOM_vs_E_vs_H}
\end{table}

\begin{table}[p]
\centering
\small
\begin{tabular}{ c|c c|c c|c c }
    \hline\hline
{energy range [keVee]} & \multicolumn{2}{c|}{fast component} & \multicolumn{2}{c|}{medium component} & \multicolumn{2}{c}{slow component} \\
350~--~1250 & decay time {[ns]} & intensity & decay time {[ns]} & intensity & decay time {[ns]} & intensity \\
    \hline
OGS gamma & 1.95~\textpm~0.19 & 85\% & 10.0~\textpm~3.7 & 9\% & 73~\textpm~14 & 6\% \\
OGS neutron & 2.32~\textpm~0.16 & 70\% & 24.2~\textpm~5.0 & 17\% & 132~\textpm~15 & 13\% \\
stilbene gamma & 4.98~\textpm~0.20 & 89\% & 23~\textpm~14 & 6\% & 107~\textpm~42 & 5\% \\
stilbene neutron & 5.88~\textpm~0.22 & 60\% & 40.5~\textpm~7.0 & 19\% & 291~\textpm~38 & 21\% \\
    \hline\hline
\end{tabular}
\caption{Pulse components decay times and intensities. Note: uncertainties of decay times were estimated as 2$\sigma$ of Gaussian fit to parameter values distribution.}
\label{tab:pulse_components}
\end{table}

\section{Conclusions} \label{sec:conclusions}
Organic glass scintillators are effective detectors of neutron and gamma radiation across a wide energy range. Their FOM parameters are comparable to those of the liquid scintillator EJ-309, provided they are optimally configured (short and long gates in the Charge Comparison Method). However, OGS scintillators exhibit stronger light self-absorption than EJ-276 plastic, which limits their maximum size. Despite this, their excellent neutron-gamma discrimination capability (FOM > 1.0) significantly surpasses that of EJ-276 scintillators.
Additionally, the short pulses of OGS offer an advantage over trans-stilbene, as they allow for detection at higher count rates.
\par
Short and long gates should be tuned depending on desired energy range of neutrons and gamma rays to be measured. This is true for both OGS and \mbox{trans-stilbene}, however the latter is more forgiving since its FOM values are generally much higher.
For measurements in a wide energy range it is advisable to choose gates values corresponding to the lowest energy. These values will minimize risk of critical failure (i.e. when neutron and gamma ray pulses cannot be distinguished) at low energy, while retaining good discrimination capability at higher energies.
\par
The optimal gates do not depend on size of the scintillator. They can be considered a material property within a certain energy range, as long as the rest of the measurement setup does not change.
\par
While pulse shape discrimination parameters vary with energy, we found no evidence that the pulse shape itself undergoes any change.

\section*{Acknowledgments}
This work was supported in part by the European Union (ChETEC-INFRA, project no. 101008324).

\bibliographystyle{unsrtnat}
\bibliography{bibliography}

\end{document}